\documentclass[aps,twocolumn,groupedaddress]{revtex4}

\begin{document}
\draft
\title{Understanding Kondo Peak Splitting and the Mechanism of Coherent Transport in a Single-Electron Transistor}

\author{Jongbae Hong and Wonmyung Woo}
\affiliation{Department of Physics and Astronomy \& Center for
Theoretical Physics, Seoul National University, Seoul 151-747,
Korea}

\date{\today}

\begin{abstract}
The peculiar behavior of Kondo peak splitting under a magnetic
field and bias can be explained by calculating the nonequilibrium
retarded Green's function via the nonperturbative dynamical theory
(NDT). In the NDT, the application of a lead-dot-lead system
reveals that new resonant tunneling levels are activated near the
Fermi level and the conventional Kondo peak at the Fermi level
diminishes when a bias is applied. Magnetic field causes asymmetry
in the spectral density and transforms the new resonant peak into
a major peak whose behavior explains all the features of the
nonequilibrium Kondo phenomenon. We also show the mechanism of
coherent transport through the new resonant tunneling level.

\end{abstract}


\maketitle \narrowtext

After the observation of the equilibrium Kondo phenomenon in a
single-electron transistor (SET)\cite{1}, nonequilibrium Kondo
phenomenon has rapidly evolved into one of the highly debated
subjects in condensed matter physics. As the phenomenon entails
two theoretically challenging field of study, namely
nonequilibrium and strong correlation, thus far no theoretical
study has been successfully able to explain the experiments
exhibiting the nonequilibrium Kondo phenomenon fully\cite{2,3}.
Therefore, a theoretical understanding of this phenomenon would be
an essential development in the advancement of condensed matter
physics. The most attractive aspect of the nonequilibrium Kondo
phenomenon is the splitting of the Kondo peak under a magnetic
field\cite{2,3}. According to the experiment performed by  Amasha
{\it et al}.\cite{3}, nonequilibrium Kondo phenomena can be
summarized as follows: (i) splitting vs. magnetic field, which is
expressed by a simple relation $\Delta_{{\rm K},S}=\Delta_{{\rm
K},S}^0+|g|\mu_BB$, where $\Delta_{{\rm K},S}$ denotes half of the
gap between the split Kondo peaks, the superscript $0$ denotes
field-independence, and the subscript $S$ denotes particle-hole
symmetric case; (ii) splitting vs. gate voltage, which is given by
$\Delta_{\rm K}^0=\Delta_{{\rm K},S}^0-C_0(V_{\rm g}-V_{{\rm
g},0})^2$, where $V_{{\rm g},0}$ denotes the gate voltage in the
middle of the Coulomb valley and the curvature of the parabola
$C_0$ is field-independent; (iii) splitting vs. Kondo temperature,
which appears to be logarithmically decreasing, i.e., $\Delta_{\rm
K}^0(T_{\rm K})\propto -C_1\ln T_{\rm K}$, where $C_1$ is also
field-independent; and (iv) the maximum Kondo temperature $T_{\rm
Km}$ vs. the critical magnetic field $B_{\rm C}$ at the splitting
threshold, which is nonlinear. One notable issue common to all the
abovementioned features is the field-independent behaviors of
$\Delta_{{\rm K},S}^0$, $C_0$, and $C_1$. Existing
theories\cite{4} cannot explain the field-independent behaviors in
(i), (ii), and (iii) and the nonlinear behavior in (iv). We will
show that these can be explained by the nonperturbative dynamical
theory (NDT)\cite{5}.

\begin{figure}[t]
\vspace*{3cm} \includegraphics{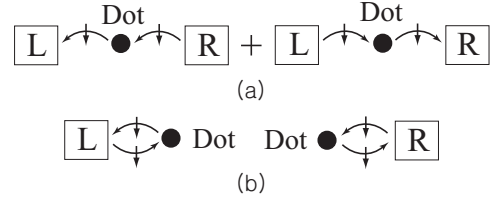} \vspace*{0.0cm} \caption{Motions of
spin-down electron in the first (a) and second (b) type of
coherence. In (b), left one corresponds to $Z^{LL}$ while right
one to $Z^{RR}$.}
\end{figure}

\begin{figure}[b]
\vspace*{4.5cm} \includegraphics{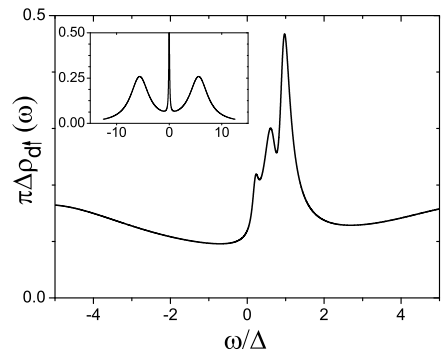} \vspace*{0cm} \caption{The
spectral density of the SET for $U=10\Delta$ under a bias, where
$\Delta=\Gamma/2$. The Zeeman shift is $0.5\Delta$. The inset is
for that under equilibrium.}
\end{figure}

The result of NDT shows that the Kondo peak in equilibrium
comprises two different types of coherence whose spectral weights
will be denoted by $Z_S^{LR}$ and $Z_S^{LL}$ below. In the first
type of coherence, the spin-down electron, for instance, travels
back and forth through the dot as shown in Fig. 1 (a), while it
moves as that shown in Fig. 1 (b) in the second type of coherence.
The spin-up electron arrives toward the dot from the left or right
lead in both the cases. The spectral density has a single resonant
peak in equilibrium as shown in the inset of Fig. 2. Because of
two separate metallic reservoirs, the maximum of
$\pi\Delta\rho_{d\uparrow}(\omega)$ is $0.5$ instead of unity.

The most interesting result of NDT appears when a bias is applied.
A part of the spectral weight of $Z_S^{LR}$ transfers from the
Fermi level to the new resonant tunneling levels near the Fermi
level, while the spectral weight $Z_S^{LL}$, which corresponds to
the second type of coherence, remains at the Fermi level. However,
the weight $Z_S^{LL}$ at the Fermi level is suppressed by
decoherence due to bias. Interesting point is that the position of
the new resonant level is independent of the applied field. This
field-independence is responsible for the field independent
features of the Kondo-peak splitting mentioned above.

When magnetic field is applied, an asymmetry occurs in the
spectral density in addition to the usual Zeeman shift
$|g|\mu_BB$. As a result of asymmetry, one of the new resonant
peaks becomes a major peak as shown in Fig. 2. All the
abovementioned features of the Kondo peak splitting phenomenon can
be explained by the position of this major peak. In order to draw
the spectral density $\rho_{d\uparrow}(\omega)$, we artificially
choose the parameters of the spectral density to show the
asymmetry and the suppression of the central peak explicitly. The
correct values of the parameters may be determined by the
self-consistent calculation proposed in Ref. [5].

The result of NDT gives the positions of the new resonant peak as
$\pm\sqrt{Z_S^{LL}}U/2$ in the Kondo regime with particle-hole
symmetry, where $U$ is the amount of Coulomb repulsion at the dot.
Therefore, the field-independent splitting in feature (i)
mentioned earlier is given by $\Delta_{{\rm
K},S}^0=\sqrt{Z_S^{LL}}U/2$. Since asymmetry is involved in the
part that becomes $U$ in the symmetric case, we propose the
following expression of $\Delta_{\rm K}^0$ for the asymmetric
case:
\begin{equation}\Delta_{\rm K}^0=\sqrt{Z_S^{LL}}
\left[\frac{4\sqrt{2}\Gamma}{\pi}
\ln\frac{\widetilde{D}}{Z\Gamma}\right],\end{equation} where
$Z=Z^{LL}+Z^{LR}$ and $\widetilde{D}=Z_S\Gamma\,{\rm exp}[\pi
U/8\sqrt{2}\Gamma]$. Equation (1) recovers $\Delta_{{\rm K},S}^0$
if $Z=Z_S$. If we use the notations given in Ref. [3], the
wavefunction renormalization $Z$ can be rewritten as $Z=Z_S{\rm
exp}[\chi(V_{\rm g}-V_{{\rm g},0})^2]=Z_ST_{\rm K}/T_{\rm K,0}$,
where $V_{\rm g}$ denotes the gate voltage and $T_{\rm K,0}$ is
the Kondo temperature at gate voltage in the middle of the Coulomb
valley $V_{{\rm g},0}$. Then, the asymmetric behavior of
$\Delta_{\rm K}^0$ is given by the parabolic or logarithmic form
\begin{eqnarray*}
\Delta_{\rm K}^0&=&\Delta_{{\rm K},S}^0-\frac{8\sqrt{2}\Gamma}{\pi
U}\Delta_{{\rm K},S}^0 \, \chi(V_{\rm
g}-V_{{\rm g},0})^2 \nonumber \\
&=&\Delta_{{\rm K},S}^0-\frac{8\sqrt{2}\Gamma}{\pi U}\Delta_{{\rm
K},S}^0\ln (T_{\rm K}/T_{\rm K,0}).\end{eqnarray*} These
expressions simultaneously explain all the abovementioned features
of the Kondo peak splitting in a qualitative manner, except for
the feature of the splitting threshold. The curvature of the
parabola and the coefficient in the $-\ln T_{\rm K}$-dependence
are given by $C_0=-8\sqrt{2}\Gamma\Delta_{{\rm K},S}^0\chi/\pi U$
and $C_1=8\sqrt{2}\Gamma\Delta_{{\rm K},S}^0/\pi U$, respectively.
Since the values of the constants $C_0$ and $C_1$ are given by
$C_0=-0.22\mu$eV/(mV)$^2$ and $C_1=11\mu$eV, respectively, using
the experimental values $\Gamma=330\mu$eV, $U=1.2$meV,
$\chi=0.020$(mV)$^{-2}$, and $\Delta_{\rm K,S}^0=11\mu{\rm eV}$,
quantitative agreement with the experimental data is perfect.
\begin{figure}[b]
\vspace*{4cm} \includegraphics{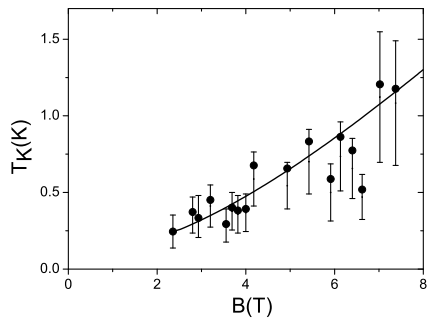} \vspace*{0cm}
\caption{Relationship between $T_{\rm Km}$ and $B_{\rm C}$ at the
splitting threshold.  }
\end{figure}

The threshold of the Kondo peak splitting depends on both the
width of the peak and the positions of the major peaks produced by
both the spin-up and spin-down electrons. Since the width of Kondo
peak is linearly proportional to the Kondo temperature, the linear
deviation of the Kondo temperature from its minimum, i.e., $T_{\rm
K}-T_{\rm K,0}$ will play an important role in constructing the
threshold equation. We now consider relative positions of the
major peaks. For the field $|g|\mu_BB=\Delta_{\rm K}^0$, the major
peak of $\rho_{d\uparrow}(\omega)$ appears at $2\Delta_{\rm K}^0$
above the Fermi level, while that of $\rho_{d\downarrow}(\omega)$
is located at the Fermi level. In this case, no separation will be
observed because the major peak of $\rho_{d\downarrow}(\omega)$
for a reversed bias has the same position with that of the forward
bias. The spectral densities for forward and reverse bias are
mirror-reflected with respect to the vertical axis.

We assume that the separation may be clearly observed when the
major peak of $\rho_{d\downarrow}(\omega)$ appears at a position
lower than $\Delta_{\rm K}^0$ below the Fermi level, i.e.,
$|g|\mu_BB+\Delta^0_{\rm K}(T_{\rm K})\geq 3\Delta^0_{\rm
K}(T_{{\rm K},0})$. If we combine this with the effect of Kondo
temperature mentioned above, we write the threshold equation as
$|g|\mu_BB_{\rm C}+\Delta^0_{\rm K}(T_{\rm Km})=3\Delta^0_{{\rm
K},S}[1+(T_{\rm Km}-T_{\rm K,0})/4T_{\rm K,0}]$, where
$\Delta^0_{{\rm K},S}=\Delta^0_{\rm K}(T_{{\rm K},0})$ and $T_{\rm
Km}$ is the maximum Kondo temperature at threshold. At $T_{\rm
K}=T_{\rm K,0}$, $B_{\rm C}=2\Delta_{{\rm K},S}^0/|g|\mu_B$, which
corresponds to $B_{\rm C}=2.4{\rm T}$ for $\Delta_{{\rm
K},S}^0=11\mu{\rm eV}$ and $|g|=0.16$\cite{3}. The coefficient
$3\Delta^0_{{\rm K},S}/4T_{\rm K,0}$ has been introduced
phenomenologically. Remarkable agreement with the experiment is
shown in Fig. 3, if we use the experimental value $T_{\rm
K,0}=0.25$K\cite{3}.

We successfully explained the experiments using Eq. (1). Now, it
is right time to derive Eq. (1)  by calculating the retarded
Green's function $G_{dd\uparrow}^+(\omega)$ under nonequilibrium
conditions. We use the NDT that provides a new technique to
calculate $G_{dd\uparrow}^+(\omega)$\cite{5}, which is expressed
as $G_{dd\sigma}^+(\omega)=\langle c_{d\sigma}|(\omega
+i\eta-{\rm\bf L})^{-1}|c_{d\sigma}\rangle$ in the Heisenberg
picture\cite{6}, where ${\rm\bf L}$ is the Liouville operator
defined by ${\rm\bf L}{A}={H}{A}-{A}{H}$ where $H$ is the
Hamiltonian, $\eta$ is a positive infinitesimal, and the inner
product is defined as $\langle A|B\rangle\equiv \langle\{{
A},{B}^\dagger\}\rangle$, where ${A}$ and ${B}$ are the operators
of the Liouville space, the curly brackets denote the
anticommutator, and the last angular brackets represent a
nonequilibrium average\cite{5}. If we define the elements of the
matrix ${\rm\bf M}$ as ${\rm\bf M}_{ij}=\langle{\hat e}_j|z{\rm\bf
I}+i{\rm\bf L}|{\hat e}_i\rangle$, where $z=-i\omega+\eta$ and the
operator ${\hat e}_j$ is one of the bases spanning the Liouville
space, the retarded Green's function is represented by
$iG_{dd\sigma}^+(\omega)=\langle c_{d\sigma}|{\rm\bf M}
^{-1}|c_{d\sigma}\rangle=({\rm adj}\,\, {\rm\bf M})_{dd}[{\rm det}
\,\,{\rm\bf M}]^{-1}$, where $({\rm adj}\,\, {\rm\bf M})_{ij}$
denotes the cofactor of the $ji$-element in the determinant of
${\rm\bf M}$.

The NDT is initiated by constructing a complete set of dynamical
bases $\{{\hat e}_j|j=1, 2, \cdots, \infty\}$ spanning the reduced
Liouville space in which the dynamics of the operator
$c_{d\sigma}(t)$ effectively describes the Kondo processes. We
simply extend the bases for the single-impurity Anderson model
used in the previous study\cite{5} to the case of two metallic
reservoirs described by the Hamiltonian
$H=\sum_{k,\sigma,\alpha=L,R}\epsilon_kc^{\alpha\dagger}_{k\sigma}
c^{\alpha}_{k\sigma}+\sum_{k,\sigma,\alpha=L,R}(V_{kd}c^\dagger
_{d\sigma}c^\alpha_{k\sigma}+V^*_{kd}c^{\alpha\dagger}_{k\sigma}c_{d\sigma})+
\sum_{\sigma=\pm 1}(\epsilon_d-\sigma g|\mu_B|B)c^\dagger
_{d\sigma}c_{d\sigma}+Un_{d\uparrow}n_{d\downarrow},$ in which $L$
and $R$ denote the left and right leads, respectively, and the
subscript $k$ denotes the quantum state of the metallic leads and
$n_{d\downarrow}=c_{d\downarrow}^{\dagger} c_{d\downarrow}$. Then,
the basis operators comprise the following five sets: (a)
$S^L_k\equiv\{c^L_{k\uparrow}|k=1, 2, \cdots, \infty\}$ for
describing the movements in the left noninteracting metallic lead,
(b) $S^L_n\equiv \{\delta n_{d\downarrow}c^L_{k \uparrow}|k=1, 2,
\cdots, \infty\}$ for describing the annihilation of a spin-up
electron in the left lead combined with the density fluctuations
in the spin-down electron at the dot, (c) and (d) the sets similar
to (a) and (b) for the right lead, i.e., $S^R_k$ and $S^R_n$,
respectively, and (e) $S_d\equiv\{\delta
j^{-L}_{d\downarrow}c_{d\uparrow}, \delta
j^{+L}_{d\downarrow}c_{d\uparrow}, c_{d\uparrow}, \delta
j^{+R}_{d\downarrow}c_{d\uparrow},
j^{-R}_{d\downarrow}c_{d\uparrow}\}$ for describing the dynamical
Kondo processes at the dot.

If we symmetrically arrange the bases such as $S_k^L$, $S_n^L$,
$S_d$, $S_n^R$, and $S_k^R$ to construct the matrix ${\rm\bf M}$,
the following matrix of nine blocks is obtained:
\[ {\rm\bf
M}_{\ell d\ell}=\left(
\begin{array}{ccc} {\rm\bf M}_{LL} \,\,\,\,\, {\rm\bf M}_{dL} \,\,
\,\, \,\,{\rm\bf 0}\,\,\,\,\,\,\, \\ {\rm\bf M}_{Ld} \,\,\,\,\,\,
{\rm\bf M}_{d} \,\,\,\,\,\,\, {\rm\bf M}_{Rd}
\\\,\,\,\, {\rm\bf 0} \,\,\,\,\,\,\,\,\,\,\,\, {\rm\bf M}_{dR}
\,\,\,\, {\rm\bf M}_{RR}
\end{array} \right),\] where the blocks ${\rm\bf M}_{d}$,
${\rm\bf M}_{dL}$ and ${\rm\bf M}_{dR}$, and ${\rm\bf M}_{Ld}$ and
${\rm\bf M}_{Rd}$ are $5\times 5$, $5\times\infty$, and
$\infty\times 5$ matrices, respectively. Blocks ${\rm\bf M}_{LL}$
and ${\rm\bf M}_{RR}$ are the $\infty\times\infty$ matrices that
are constructed by the sets of bases describing the left and right
leads, respectively. Since no direct coupling exists between the
left and right leads, zero matrices occur at the two corners. The
structure of each block is similar to that of the corresponding
block of the matrix for the Anderson model considered in the
previous study\cite{5}.

The infinite-dimensional matrix can be reduced to a
finite-dimensional matrix using the L\"{o}wdin's partitioning
technique\cite{7,8}. It is performed by solving the eigenvalue
equation for the matrix ${\rm\bf M}_{\ell d\ell}$, such as
${\rm\bf M}_{\ell d\ell}{\rm\bf C}={\rm\bf 0}$, where ${\rm\bf C}$
and ${\rm\bf 0}$ are the infinite-dimensional column vectors. The
column vector ${\rm\bf C}$ is partitioned into three parts, i.e.,
${\rm\bf C}^T=({\rm\bf C}^L \, {\rm\bf C}^d \, {\rm\bf C}^R)$,
where $T$, $L$, $d$, and $R$ denote the transpose, left lead, dot,
and right lead, respectively. Then, the equation for ${\rm\bf
C}^{d}$ is obtained as $({\rm\bf M}_{d}-{\rm\bf M}_{Ld}{\rm\bf
M}_{LL}^{-1}{\rm\bf M}_{dL}-{\rm\bf M}_{Rd}{\rm\bf
M}_{RR}^{-1}{\rm\bf M}_{dR}){\rm\bf C}^{d}\equiv {\widetilde
{\rm\bf M}}_{d}{\rm\bf C}^{d}={\rm\bf 0}$. The reduced $5\times 5$
matrix ${\widetilde{\rm\bf M}}_{d}$ contains information on the
many-body dynamics of the spin-up electron. Obtaining ${\widetilde
{\rm\bf M}}_{d}$ is practically possible if the
$\infty\times\infty$ matrices ${\rm\bf M}_{LL}$ and ${\rm\bf
M}_{RR}$ are block diagonal\cite{9}. The second and third terms
appear as self-energies in ${\widetilde{\rm\bf M}}_{d}$ after
reduction.

The matrix ${\widetilde{\rm\bf M}}_{d}$ is expressed as
\[ \widetilde{\rm\bf M}_{d}=\left( \begin{array}{ccccc} -i\tilde{\omega}
\,\,\,\,\,\,\,  -\gamma_{LL} \,\,\,\, \,\, \,\, -U^L_{J^-}
\,\,\,\, \,\, \,\, \gamma_{LR} \,\,\,\, \,\, \,\, \gamma_{J^-} \\
\\ \,\,  \gamma_{LL} \,\, \,\,\,\, \,\,  -i\tilde{\omega}
\,\,\,\,\,\,
 -U^L_{J^+} \,\, \,\, \,\,\,\, \,\,\gamma_{J^+} \, \,\,\,\,  \,\, \gamma_{LR} \\ \\
\,\,\, \,\,\,U_{J^-}^{L*} \,\,\, \,\,\,\,\, \,\,U_{J^+}^{L*}
\,\,\,\,\,\,\, -i\tilde{\omega} \,\, \,\, \,\,\,\,\,\,\,
U^{R*}_{J^+} \,\,\,\, \,\, \,\,\,\, U^{R*}_{J^-}
\\ \\ \,\, -\gamma_{LR} \,\,\,\, \,\, \,\, -\gamma_{J^+} \,\,\,\,
\,\, \,\, -U_{J^+}^R  \,\,\,\, -i\tilde{\omega} \,\, \,\, \,\,\, \,\,\, \gamma_{RR} \\ \\
\,\, \,\, -\gamma_{J^-} \,\, \,\,\,\,  -\gamma_{LR} \,\, \,\,\,
 -U_{J^-}^R\,\, -\gamma_{RR} \,\,\, -i\tilde{\omega}
\end{array} \right), \]
where $\tilde{\omega}\equiv\omega-\epsilon_d-U\langle
n_{d\downarrow}\rangle+g|\mu_B|B$. All the matrix elements, other
than $U_{J^\pm}$ and $U_{J^\pm}^*$, have additional self-energy
functions
$i\Sigma_{ij}(\omega)=\beta_{ij}(i\Sigma^L_0(\omega)+i\Sigma^R_0(\omega))$,
where $\Sigma_0(\omega)$ is the self-energy for the Anderson model
at $U=0$\cite{10}. The coefficient $\beta_{33}=1$ and others,
$\beta_{ij}=\beta_{ji}$, are given by
$\beta_{22}=\beta_{24}=\beta_{44}> \beta_{12}=\beta_{14}>
\beta_{11}=\beta_{55}>\beta_{15}>\beta_{25}=\beta_{45}$\cite{11}.
In equilibrium at half-filling, however, $\beta_{ij}=1/4$ except
$\beta_{33}$. We use that $\gamma_{LL}=\gamma_{RR}$ and
$\gamma_{J^-}=\gamma_{J^+}$. The latter will be discussed below.

In particular, it is noteworthy that the zeros of the determinant
of $\widetilde{\rm\bf M}_{d}$ are $\tilde{\omega}=0,
\pm[\gamma_{LL}^2+(\gamma_{LR}-\gamma_{J^{\pm}})^2]^{1/2}$, and
$\pm U/2$ in the large-$U$ and atomic limit. The second zeros
become $\pm\gamma_{LL}$ under appropriate amount of bias. This
implies that there are additional resonant tunneling levels at
$\pm\gamma_{LL}$ if we treat the system with two separate leads.
One of the effects of the magnetic field is the Zeeman shift
appearing in $\tilde{\omega}$ in the diagonal elements. We use
$\omega$ to represent $\omega-\epsilon_d-U\langle
n_{d\downarrow}\rangle$ hereafter. The Zeeman shift results in a
nonvanishing value of $1/2-\langle n_{d\downarrow}\rangle$ that
induces asymmetry in the spectral density by the nonvanishing
imaginary part of the elements $U_{J^\pm}$, which can be expressed
as $U_{J^\pm}=(U/4)[1+i2(1-2\langle n_{d\downarrow}\rangle)\langle
j^\pm_{d\downarrow}\rangle/\langle(\delta
j^\pm_{d\downarrow})^2\rangle^{1/2}]$. The asymmetry maximizes the
new resonant peak appearing at $\omega=g\mu_BB+\gamma_{LL}$, as
shown in Fig. 2. Therefore, the field-independent part of the
splitting $\Delta^0_{{\rm K},S}$ is merely $\gamma_{LL}$. The real
part of $U_{J^\pm}$ determines the positions of the $U$-peaks. The
coefficient $1/4$ was determined from the analysis performed at
the atomic limit. It is $1/\sqrt{2}$ times smaller than that of
the single impurity Anderson model\cite{5}.

Since $\gamma_{LL}=\langle \sum_k
i(V_{kd}^{*L}c_{k\uparrow}^L+V_{kd}^{*R}c_{k\uparrow}^R)
c^\dagger_{d\uparrow}[j^{-L}_{d\downarrow},j^{+L}_{d\downarrow}]\rangle$,
its direct calculation is difficult\cite{12}. We have tried direct
calculation of $\gamma_{LL}$ in our previous work\cite{13}. We
have obtained that $\gamma_{LL}\propto\ln
(\widetilde{D}/Z\Gamma)$. However, we failed to derive a correct
prefactor of $\ln (\widetilde{D}/Z\Gamma)$ in the previous work,
which is necessary to explain the experimental results
quantitatively. Even though the NDT provides us all qualitative
features of the Kondo peak splitting phenomenon, we need correct
prefactor of $\ln (\widetilde{D}/Z\Gamma)$ for the quantitative
comparison with experiment. An indirect way employing the exact
result by Bethe Ansatz\cite{10} make it possible to derive an
appropriate expression of $\gamma_{LL}$.

The wavefunction renormalization $Z$ can also be obtained by
calculating the spectral weight of $\rho_{d\uparrow}(\omega)$ at
$\omega=0$. Under equilibrium with particle-hole symmetry, it is
given by $Z_S=[1+U^2/\{4(\gamma_{LL}^2+\gamma_{LR}^2)\}]^{-1}$,
where $\gamma_{LR}=\langle\sum_ki
(V_{kd}^{*L}c_{k\uparrow}^L+V_{kd}^{*R}c_{k\uparrow}^R)c^\dagger_{d\uparrow}
[j^{-L}_{d\downarrow},j^{+R}_{d\downarrow}]\rangle$. In the Kondo
regime, $Z_S$ gets separated into two parts, i.e.,
$Z^{LL}_S=4\gamma_{LL}^2/U^2$ and $Z^{LR}_S=4\gamma_{LR}^2/U^2$.
The former is the spectral weight remaining at the Fermi level,
which does not contribute to the Kondo peak splitting, and the
latter, which is responsible for the Kondo peak splitting,
transfers to the new resonant tunneling level at
$\omega=\gamma_{LL}$ and gives rise to the additional shift
$\sqrt{Z_S^{LL}}U/2$ to the one by Zeeman effect.
\begin{figure}[t]
\vspace*{5.5cm} \includegraphics{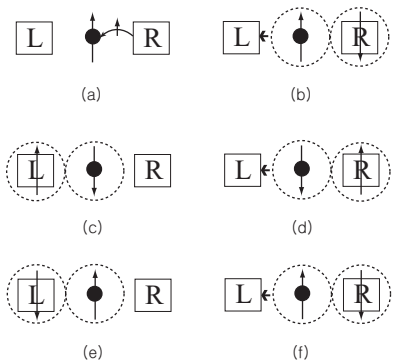} \vspace*{0cm} \caption{Coherent
transport through the new resonant tunneling level when a spin-up
electron hops into the dot. }
\end{figure}

As one can see from the operator expression of $\gamma_{LL}$, it
is independent of the Coulomb repulsion. Therefore, $U$ appearing
in the expression of $\gamma_{LL}$ should come from the averaging
process and may be expressed by $Z$. The theory using Bethe ansatz
yields $Z_S=(4/\pi)\sqrt{U/\Gamma}{\rm exp}[-\pi U/4\Gamma]$ for
the symmetric Anderson model in the Kondo regime\cite{10}. Then,
$U$ is written as $U=(4\Gamma/\pi)\ln(\widetilde{D}/Z_S\Gamma)$,
where $\widetilde{D}=Z_S\Gamma{\rm exp}[\pi U/4\Gamma]$. For the
Anderson model with two separate reservoirs, however, the effect
of the two reservoirs changes $U$ and $\Gamma$ into $U/\sqrt{2}$
and $2\Gamma$, respectively. Therefore,
$Z_S=(4/\pi)\sqrt{U/2\sqrt{2}\Gamma}{\rm exp}[-\pi
U/8\sqrt{2}\Gamma]$ for the SET in the Kondo regime. By using the
expression for $U$ from the $Z_S$ for the SET and removing the
subscript $S$ to take the asymmetry into account, we obtain
$\gamma_{LL}=(4\sqrt{2}\Gamma/\pi)\sqrt{Z_S^{LL}}\ln
(\widetilde{D}/Z\Gamma)$, where $\widetilde{D}=Z_S\Gamma{\rm
exp}[\pi U/8\sqrt{2}\Gamma]$. This is just Eq. (1).

It is interesting and meaningful to observe the mechanism of
electron transport through the new resonant tunneling level in an
SET. This mechanism appears in the operator expressions of
$\gamma_{LR}$. The motion of the spin-down electron is governed by
the operator $[j^{-L}_{d\downarrow},j^{+R}_{d\downarrow}]$, as
described in Fig. 1 (a), while spin-up electron moves into the dot
from the left or the right lead.  Figure 4 shows the process
making current when the chemical potential of the right lead is
higher than that of left. Since the movement that costs energy $U$
is not allowed, a possible operation that constitutes a current is
the one in which a spin-up electron moves from the right lead to
the dot, makes a singlet coupling with a spin-down electron on the
right lead, and moves together toward the left lead. Then, the
spin-down electron changes its partner that is a spin-up electron
on the right lead to make a singlet. This new singlet moves to the
left and the process is repeated. It is noteworthy that the
singlet state is retained during transport and the partner of
singlet is changed, similar to that in superconductivity.

The motion of the spin-down electron of the operator $[j^{\mp
L}_{d\downarrow},j^{\mp R}_{d\downarrow}]$ in
$\gamma_{J^\mp}=\langle\sum_ki
(V_{kd}^{*L}c_{k\uparrow}^L+V_{kd}^{*R}c_{k\uparrow}^R)
c^\dagger_{d\uparrow}[j^{\mp L}_{d\downarrow},j^{\mp
R}_{d\downarrow}]\rangle$ is Fig. 1 (a) with $-$ sign, which
corresponds to the net flow of the spin-down electrons. When the
bias increases, $\gamma_{J^\mp}$ approaches $\gamma_{LR}$ and the
spectral weight of the new resonant peak also increases. However,
at equilibrium, $\gamma_{J^\mp}$ vanishes and the new resonant
peak disappears.

In conclusion, we report the existence of a new resonant tunneling
level in an SET with a strong correlation. This new resonant
tunneling level is activated only under nonequilibrium conditions,
and it explains all the features of the Kondo peak splitting
phenomenon  measured by Amasha {\it et al}.\cite{3}. Further, we
show in Fig. 4 the coherent transport mechanism through the dot
using the new resonant tunneling level.

This work was supported by the Korea Research Foundation, Grant
No. KRF-2006-0409-0060.


\begin{references}
\bibitem{11} The expression $\beta_{12}$, for instance, is given by
$\beta_0[1/4-A/[\sigma_r(
j^{+L}_{d\downarrow})\sigma_r(j^{-L}_{d\downarrow})]$, where
$\beta_0=1/4\langle(\delta n_{d\downarrow})^2\rangle$,
$A=(1-2\langle n_{d\downarrow}\rangle)^2$, and
$\sigma_r(j_{d\downarrow}^{\pm})=\langle(\delta
j_{d\downarrow}^{\pm})^2\rangle^{1/2}/\langle
j_{d\downarrow}^{\pm}\rangle$.
\bibitem{12} We omit the normalization factor in the expressions of all $\gamma$'s.
\end{references}
\end{document}